\documentstyle[prd,aps,preprint]{revtex}

\newcommand{\bdi}{\begin{displaymath}}
\newcommand{\edi}{\end{displaymath}}
\newcommand{\bfi}{\begin{figure}}
\newcommand{\efi}{\end{figure}}

\newcommand{\beq}{\begin{equation}}
\newcommand{\eeq}{\end{equation}}

\newcommand{\gaf}{\gamma_{5}}

\newcommand{\beqa}{\begin{eqnarray}}
\newcommand{\eeqa}{\end{eqnarray}}
\newcommand{\no}{\nonumber}
\newcommand{\ra}{\rightarrow}

\newcommand{\wt}{\widetilde}

\newcommand{\dsla}{\partial\hspace{-6pt} /  }  
\newcommand{\Asla}{A\hspace{-6.5pt}  /  }

\begin{document}

\setcounter{page}{0}
\def\footnoterule{\kern-3pt \hrule width\hsize \kern3pt}
\tighten
\title{SCATTERING PROCESSES IN THE MASSIVE SCHWINGER MODEL\thanks
{This work is supported by a Schr\"odinger Stipendium of the Austrian FWF and
in part by funds provided by the U.S.
Department of Energy (D.O.E.) under cooperative 
research agreement \#DF-FC02-94ER40818.}}

\author{Christoph Adam\footnote{Email address: {\tt adam@ctp.mit.edu, 
adam@pap.univie.ac.at}}}

\address{Center for Theoretical Physics \\
Laboratory for Nuclear Science \\
and Department of Physics \\
Massachusetts Institute of Technology \\
Cambridge, Massachusetts 02139 \\
and \\
Inst. f. theoret. Physik d. Uni Wien \\
Boltzmanngasse 5, 1090 Wien, Austria \\
{~}}

\date{MIT-CTP-2602,~ hep-th/9701013. {~~~~~} January 1997}
\maketitle
\thispagestyle{empty}

\begin{abstract}

We derive the (matrix-valued) Feynman rules of the mass perturbation theory
and use it for the resummation of the $n$-point functions with the help of
the Dyson-Schwinger equations.
We use these results for a short review of the complete spectrum of the
model and for a discussion of scattering processes.
We find that in scattering cross sections all the resonances and higher
particle production thresholds of the model are properly taken into account
by our resummed mass perturbation theory, without the need of further 
approximations.

\medskip

PACS-Numbers:   11.10, 11.80, 11.90

\end{abstract}
\vspace*{\fill}


\section{Introduction}

\input psbox.tex
\let\fillinggrid=\relax

The massive Schwinger model is QED$_2$ with one massive fermion,
\beq
L=\bar\Psi (i\dsla -e\Asla +m)\Psi -\frac{1}{4}F_{\mu\nu}F^{\mu\nu}
\eeq
Both massive and massless ($m=0$) Schwinger model have been subject to 
intensive study as simple models that nevertheless show some
nontrivial field theoretic features, like anomalies, nontrivial vacuum 
structure ($\theta$ vacuum), fermion condensates, etc. 
(\cite{Sc1} -- \cite{Sm3}).
Further, they have been used as test labors for the study of some concepts
that are important for more realistic models like QCD (confinement, OPE,
etc., \cite{AAR}, \cite{GKMS} -- \cite{NSVZ}).

The massless model may be solved exactly and is, in fact, equivalent to
the theory of one free, massive boson (``Schwinger boson'') with mass
$\mu_0^2 =\frac{e^2}{\pi}$ (here the nontrivial vacuum structure may be 
detected by chiral VEVs, leading e.g. to the fermion condensate).

In the massive model the Schwinger boson turns into an interacting particle
with renormalized mass $\mu$, that may form bound states and undergo 
scattering. Mass perturbation theory (which may be performed because of the 
exact solubility of the massless model) turns out to be especially useful
for the discussion of these features, because it is formulated in terms of 
physical fields only (the Schwinger boson and chiral currents; the confined
fermions themselves do not occur), and because the mass perturbation expansion 
is about the physical $\theta$ vacuum. Actually, the mass perturbation theory
has some similarity with the chiral perturbation theory of QCD.

In this article we will systematically perform the mass perturbation 
expansion and derive the corresponding Feynman rules, which turn out to acquire
a matrix structure because of the chiral properties of the model (i.e.
because the mass term $m\bar\Psi\Psi$ mixes left- and right-handed fields).
Further we will show how Schwinger boson $n$-point functions may be
re-expressed in terms of chiral $n$-point functions by the use of the
Dyson-Schwinger equations of the model. The latter ones may be resummed by
further re-expressing them in terms of non-factorizable $n$-point functions.
This resummation enables us to infer all the (stable) particle and
(unstable) bound-state masses from the poles of the two-point function.
We will find, as one result, that the spectrum of the model is richer than
expected earlier (\cite{Co1,GBOUND}). Further, by rewriting the higher 
$n$-point functions in terms of non-factorizable ones, we are able to
identify all the possible final states of decays and scattering processes,
and we may compute scattering cross sections that include the effects of all
possible resonances and higher particle production thresholds. 

All computations are based on the Euclidean path integral formalism, and
therefore, we have to take care of our specific Euclidean conventions
(see e.g. \cite{ABH}) in the computations.

\section{Mass perturbation theory}

First let us shortly review the mass perturbation theory. By simply expanding
the mass term, the vacuum functional and VEVs of the massive model may be
traced back to space-time integrations of VEVs of the massless model. E.g. the
vacuum functional is
\bdi
Z(m,\theta )= \sum_{k=-\infty}^\infty e^{ik\theta}N\int D\bar\Psi D\Psi
DA_\mu^k \cdot
\edi
\beq
\cdot \sum_{n=0}^\infty \frac{m^n}{n!}\prod_{i=1}^n \int dx_i \bar\Psi
(x_i)\Psi (x_i)\exp \int dx\Bigl[ \bar\Psi (i\dsla -e\Asla )\Psi -\frac{1}{4}
F_{\mu\nu}F^{\mu\nu}\Bigr]  
\eeq
($k$ \ldots instanton number).
Therefore, one needs scalar VEVs $\langle S(x_1)\ldots S(x_n)\rangle_0$ of the
massless model, where $S=\bar\Psi \Psi$, $S_\pm =\frac{1}{2}\bar\Psi (1\pm \gaf
)\Psi$. Chiral VEVs $\langle S_{H_1}\ldots S_{H_n}\rangle_0$, $H_i =\pm$, are
especially easily computed, as only a definite instanton sector contributes
(see e.g. \cite{MSSM,GBOUND,Zah}),
\beq
\langle S_{H_1}(x_1)\cdots S_{H_n}(x_n)\rangle_0 = e^{ik\theta} 
\Bigl( \frac{\Sigma}{2}\Bigr)^n \exp
\Bigl[ \sum_{i<j}(-)^{\sigma_i \sigma_j}4\pi D_{\mu_0} (x_i -x_j)\Bigr] 
\eeq
\bdi
k=\sum_{i=1}^n \sigma_i =n_+ -n_-
\edi
where $\sigma_i =\pm 1$ for $H_i =\pm $, $D_{\mu_0}$ is the massive scalar
propagator of the Schwinger boson ($\mu_0^2 =\frac{e^2}{\pi}$) 
and $\Sigma$ is the fermion condensate of the massless model.

The Schwinger boson $\Phi$ is related to the vector current, $J_\mu
=\frac{1}{\sqrt{\pi}}\epsilon_{\mu\nu}\partial^\nu \Phi$, and, therefore,
the $S$ and $\Phi$ VEVs, which we need for the perturbative calculation of
massive VEVs, are related to the vector and scalar current VEVs of the
massless model. Explicitly the $S$ and $\Phi$ VEVs may be computed from the
generating functional (which is at the same time a VEV for $n$ chiral currents)
\bdi
\langle S_{H_1}(x_1)\cdots S_{H_n}(x_n)\rangle_0 [\lambda]= e^{ik\theta} 
\Bigl( \frac{\Sigma}{2}\Bigr)^n \exp
\Bigl[ \sum_{i<j}(-)^{\sigma_i \sigma_j}4\pi D_{\mu_0} (x_i -x_j)\Bigr] \cdot 
\edi
\beq
\cdot \exp\Bigl[-\int dy_1 dy_2 \lambda (y_1) D_{\mu_0}(y_1 -y_2)\lambda (y_2)
+2i\sqrt{\pi}\sum_{l=1}^n (-)^{\sigma_l} \int dy\lambda (y)D_{\mu_0}(y-x_l)
\Bigr] 
\eeq 
(see \cite{Gatt2} for an explicit computation),
where $\lambda$ is the external source for the Schwinger boson $\Phi$. 
Observe the $(-)^{\sigma_l}$ in the last term of the exponent. As a
consequence, whenever an {\em odd} number of external $\Phi$ lines meets at a
point $x_i$, 
the corresponding $S_- (x_i)$ acquires a $-$, i.e. instead of a $S=S_+
+S_-$ vertex there is a $P=S_+ -S_-$ vertex.

From equs. (3), (4) one finds that exponentials $\exp \pm 4\pi D_{\mu_0} (x_i
-x_j)$ are running from any vertex $x_i$ to any other vertex $x_j$; however, in
order to get an IR finite perturbation theory, one has to expand these
exponentials into the functions
\beq
E_\pm (x)=e^{\pm 4\pi D_{\mu_0}(x)} -1,
\eeq
(or their Fourier transforms $\wt E_\pm (p)$ for momentum space Feynman rules).
This expansion is analogous to the cluster expansion of statistical physics.

The important point is that each vertex contains (for $\theta \neq 0$) two
types of vertices, $m\langle S(x)\rangle_0 =m(\langle S_+ (x)\rangle_0 +\langle
S_- (x)\rangle_0 )\simeq m\frac{\Sigma}{2}e^{i\theta}
+m\frac{\Sigma}{2}e^{-i\theta}$, and
these two types of vertices are connected by two types of propagators, namely
$S_+ (x) S_+ (y)$ and $S_- (x) S_- (y)$ by $E_+ (x-y)$, and $S_+ (x)S_- (y)$
and $S_- (x) S_+ (y)$ by $E_- (x-y)$. Further, because all vertices may be
connected to each other, up to $n-1$ lines $E_\pm (x-y_i)$ may run from one
vertex $x$ to the other vertices $y_i$ for a $n$-th order mass 
perturbation contribution.

As a consequence, the Feynman rules acquire a matrix structure. More precisely,
the propagator corresponding to the $E_\pm (x)$ is a matrix, which in momentum
space reads
\beq
{\cal E}(p) =\left( \begin{array}{cc}\wt E_+ (p) & \wt E_- (p) \\ \wt E_- (p)
 & \wt E_+ (p) \end{array} \right)
\eeq
where the individual entries correspond to the individual $\langle S_i (x)S_j
(y)\rangle_0$, $i,j=\pm$, propagators. 

Each vertex, where $n$ propagator lines ${\cal E}(p_i)$
meet, is a $n$-th rank tensor ${\cal
G}$. Only two components of this tensor are nonzero, namely
\beq
{\cal G}_{++\cdots +} =g_\theta \quad ,\quad {\cal G}_{--\cdots -} =g^*_\theta
\eeq
(corresponding to $S=S_+ +S_-$). E.g. the vertex where two propagators meet is
a matrix
\beq
{\cal G} =\left( \begin{array}{cc} g_\theta & 0 \\ 0
 & g_\theta^* \end{array} \right) .
\eeq
where $g_\theta ,g^*_\theta$ are the renormalized couplings including all
tadpole-like corrections (see e.g. \cite{GBOUND}),
\beq
g_\theta =m\frac{\Sigma}{2}e^{i\theta} +o(m^2).
\eeq
Internal lines must be ${\cal E}(p)$ because of the Feynman rules; external
lines, however, may be boson lines, too, when we treat bosonic $n$-point
functions $\langle \Phi (x_1)\ldots \Phi (x_n)\rangle_m$. There the rule is
that, when the external boson lines are amputated, the vertex where they meet
is multiplied by $P$ ($S$) when an odd (even) number of bosons meets at that
vertex, where
\beq
P = {1 \choose -1} \quad ,\quad S={1 \choose 1} .
\eeq
These Feynman rules may be given by the graphs of Fig. 1 (we display them 
in momentum space)

$$\psannotate{\psboxscaled{700}{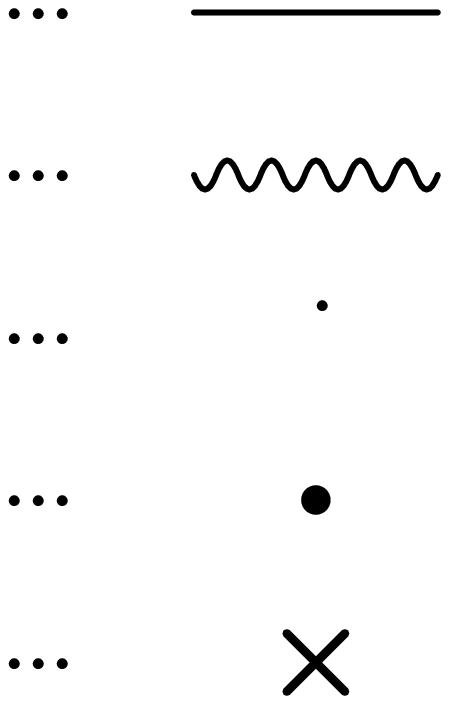}}{\fillinggrid
\at(0.7\pscm;-0.5\pscm){Fig. 1} \at(-1.9\pscm;2.8\pscm){${\cal G}$}
\at(-2.2\pscm;4.5\pscm){${\cal G}_0$} \at(-2\pscm;1.1\pscm){1}
\at(-2.6\pscm;6.2\pscm){${\cal E}(p)$} \at(-3\pscm;7.9\pscm){$\wt
D_{\mu_0}(p)$}}$$

\vspace{0.3cm}

where ${\cal G}_0$ denotes the bare coupling.
Let us e.g. investigate the connected
two-point function $\langle i\Phi (x_1) i\Phi (x_2)\rangle_m^c$
within mass perturbation theory. After amputation of the two
external boson lines, and ignoring the $P$ vectors (10) 
at the initial and final
vertices we find the following graphical representation (see Fig. 2a, b)

$$\psannotate{\psboxscaled{700}{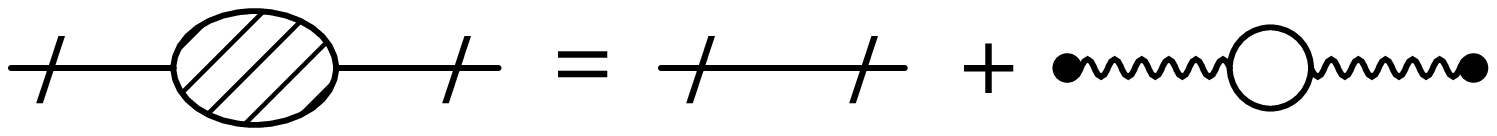}}{\fillinggrid
\at(6.9\pscm;-0.5\pscm){Fig. 2a}}$$

$$\psannotate{\psboxscaled{700}{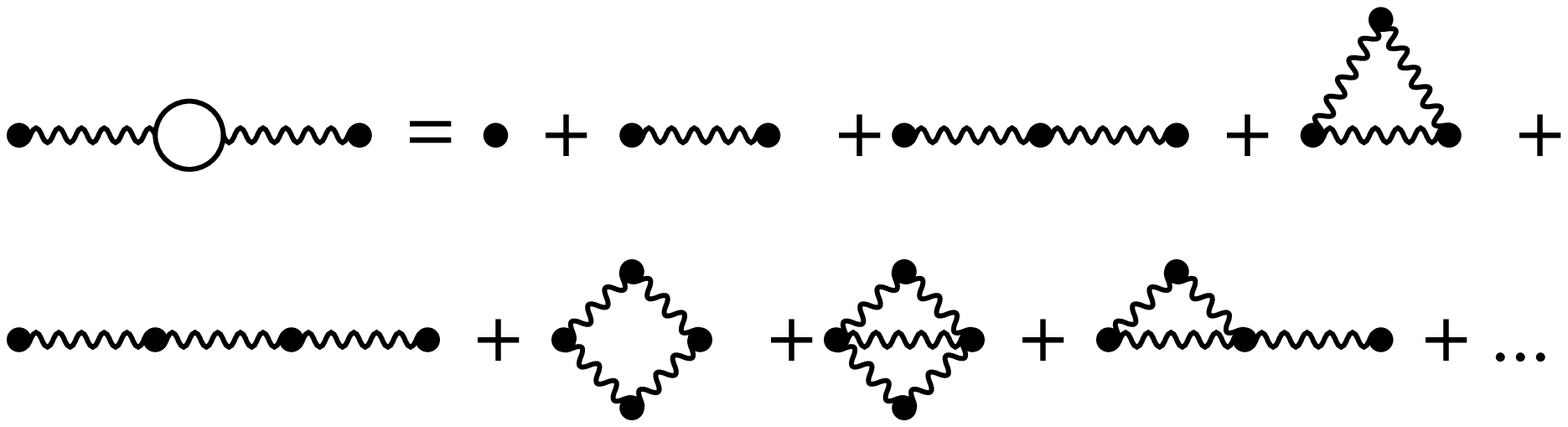}}{\fillinggrid
\at(8.9\pscm;-0.5\pscm){Fig. 2b}}$$

where always the left and right vertices are the initial and final ones.
Observe that we do not draw graphs where a sequence of wavy lines begins and
ends at the same vertex, because they are already taken into account by the
renormalized coupling ${\cal G}$. 

Introducing for the above two-point function (Fig. 2b) the name ${\cal G}\Pi
(p)$ in momentum space (matrix multiplication is understood)
\beq
{\cal G}\Pi (p) := {\cal G} +{\cal G}{\cal E}(p){\cal G}+{\cal G}{\cal E}(p)
{\cal G}{\cal E}(p){\cal G} +\ldots 
\eeq
we will find a resummation for $\Pi (p)$ that relies on the following
observation. All diagrams that fall into two pieces when they are cut at a
vertex, factorize in momentum space, i.e. they are a product of two functions
of $p$. The opposite type graphs are called non-factorizable (n.f.).

Here we should be more precise about the cutting. We stated above that the
vertices are tensors, so how to cut such a vertex? Suppose e.g. we have a
vertex where three lines meet and we want to cut it in a way that two wavy
lines belong to the left hand side, and one line to the right hand side. Then
we rewrite the vertex like
\beq
{\cal G}_{ijk}=\delta_{ijl}{\cal G}_{ll'}\delta_{l' k}\quad ,\quad
i,j,k,l,l' =\pm
\eeq
where ${\cal G}_{ll'}$ is the vertex matrix (8) and the $\delta_{i_1 \cdots
i_n}$ are generalizations of the Kronecker delta $\delta_{ij}$, i.e.
\beq
\delta_{++\cdots +}=\delta_{--\cdots -}=1\quad ,\quad \delta_{i_1 \cdots i_n}=0
\quad \mbox{otherwise}
\eeq
Therefore, we may write for the sum of non-factorizable graphs, which we
call ${\cal A}$ (see Fig. 3)

$$\psannotate{\psboxscaled{600}{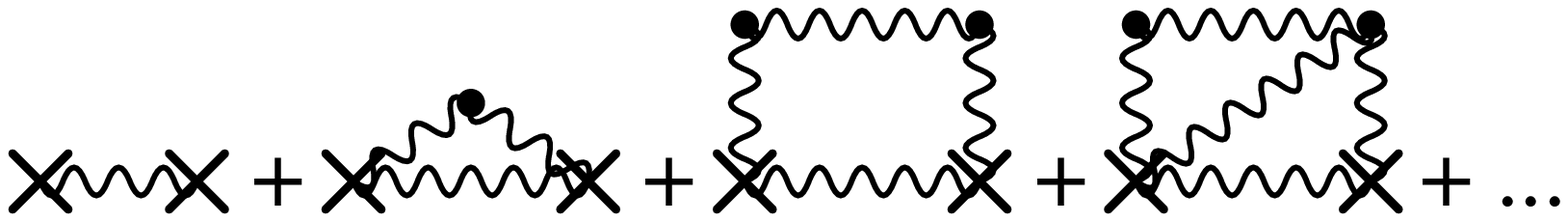}}{\fillinggrid
\at(6.9\pscm;-0.5\pscm){Fig. 3}}$$ 

\beq
{\cal A}_{ij}(p)={\cal E}_{ij}(p)+\int\frac{d^2 q}{(2\pi)^2}\delta_{ikk'}
{\cal E}_{kl}(q){\cal G}_{ll'}{\cal E}_{l' m}(q){\cal E}_{k' m'}(q-p)
\delta_{jmm'} +\ldots 
\eeq
The matrix ${\cal A}(p)$ may be rewritten like
 \beq
{\cal A}(p) =\left( \begin{array}{cc} \wt{\langle S_+ S_+ \rangle}_{\rm n.f.} (p)
 & \wt{\langle S_+ S_- \rangle}_{\rm n.f.} (p) \\ \wt{\langle S_- S_+
 \rangle}_{\rm n.f.} (p)
 & \wt{\langle S_- S_- \rangle}_{\rm n.f.} (p) \end{array} \right)
\eeq
The entries of this matrix are, however, related (e.g. $\wt{\langle S_- S_-
\rangle}_{\rm n.f.}(g_\theta ,p)=\wt{\langle S_+ S_+ \rangle}_{\rm n.f.}
(g^*_\theta ,p)$, as may be checked from the perturbative expansion) and,
therefore, we find for the product ${\cal G}{\cal A}(p)$ (which we need in the
sequel)
\beq
{\cal G}{\cal A}(p)=:\left( \begin{array}{cc} \alpha (g_\theta ,p) & \beta
(g_\theta ,p) \\ \beta (g^*_\theta ,p) & \alpha (g^*_\theta ,p) \end{array}
\right)
\eeq
\beq
\alpha (g_\theta ,p)=g_\theta \wt{\langle S_+ S_+ \rangle}_{\rm n.f.}(g_\theta
,p) \quad ,\quad \beta (g_\theta ,p)=g_\theta \wt{\langle S_+ S_- \rangle}_{\rm
n.f.}(g_\theta ,p).
\eeq
(Remark: in \cite{GBOUND} we wrote $\alpha^* (g_\theta ,p), \beta^* (g_\theta
,p)$ instead of $\alpha (g^*_\theta ,p), \beta (g^*_\theta ,p)$. This makes no
difference as long as the propagator functions $\wt E_\pm (p)$ themselves are
purely real ( or one needs the real part only). However, when the propagators
acquire imaginary parts, equ. (16) is the correct one.)

Now we may collect all n.f. graphs in (11), 
Fig. 2b, e.g. on the left hand side,
and find that they are again multiplied by {\em all} graphs that occur in Fig.
2b. Therefore we may write for ${\cal G}\Pi (p)$ of equ. (11)
\beq
{\cal G}\Pi (p)={\cal G}({\bf 1}+{\cal A}(p){\cal G}\Pi (p)).
\eeq
Equation (18) may be solved for $\Pi (p)$ by a matrix inversion and has the
solution
\beq
\Pi (p)=\frac{1}{N(p)} \left( \begin{array}{cc} 1-\alpha (g^*_\theta ,p) & 
\beta (g^*_\theta ,p) \\ \beta (g_\theta ,p) &
1-\alpha (g_\theta ,p) \end{array} \right)
\eeq
where $N(p)$ is the determinant of the matrix that had to be inverted,
\beq
N(p)=\det ({\bf 1}-{\cal G}{\cal A}(p))=1-\alpha (g_\theta ,p)-\alpha
(g^*_\theta ,p)+\alpha (g_\theta ,p)\alpha (g^*_\theta ,p) -\beta (g_\theta
,p)\beta (g^*_\theta ,p).
\eeq

\section{Dyson-Schwinger equations and higher $n$-point functions}

Observe that the relation (Fig. 2a) between the amputated Schwinger-boson 
two-point function and the graph (Fig. 2b) for ${\cal G}\Pi (p)$ is the first
Dyson-Schwinger equation of the model that may be derived from the equation of
motion
\beq
M_x i\Phi (x):=(\Box_x -\mu_0^2 )i\Phi (x)=2\sqrt{\pi}mP(x).
\eeq
Defining the Fourier transforms
\beq
M^{(n)}(p_1 ,\ldots ,p_n):={\rm FT}(M_{x_1}\ldots M_{x_n}\langle i\Phi (x_1)
\ldots i\Phi (x_n)\rangle_m^c )
\eeq
Fig. 2a may be written like ($M^{(2)}(p,p)\equiv M^{(2)}(p)$)
\beq
M^{(2)}(p)=-(p^2 +\mu_0^2 )+4\pi m\langle S\rangle_m +4\pi m^2 \wt{\langle
PP\rangle}_m^c (p)
\eeq
where
\beq
m\langle S\rangle_m \equiv P^T {\cal G}P=S_i{\cal G}_i =g_\theta +g^*_\theta
\eeq
\beq
m^2 \wt{\langle PP\rangle}_m^c (p)\equiv P^T {\cal G}(\Pi (p) -{\bf 1})P
\eeq
where $P^T =(1,-1)$ is the transpose of the vector $P$, (10), and matrix
multiplication is understood (the single vertex ${\cal G}$ we may interpret
either as a two-component object that is contracted by two vectors $P$ or as a
one-component object that is contracted by one vector $S$).

In Fig. 2a we did not display the $4\pi$ factors of (23), and we will continue
not to display them in the figures (however, we, of course, retain them in the
formulae).

Analogously one may find higher Dyson-Schwinger equations. Before showing them
we need some more graphical rules (see Fig. 4)

$$\psannotate{\psboxscaled{700}{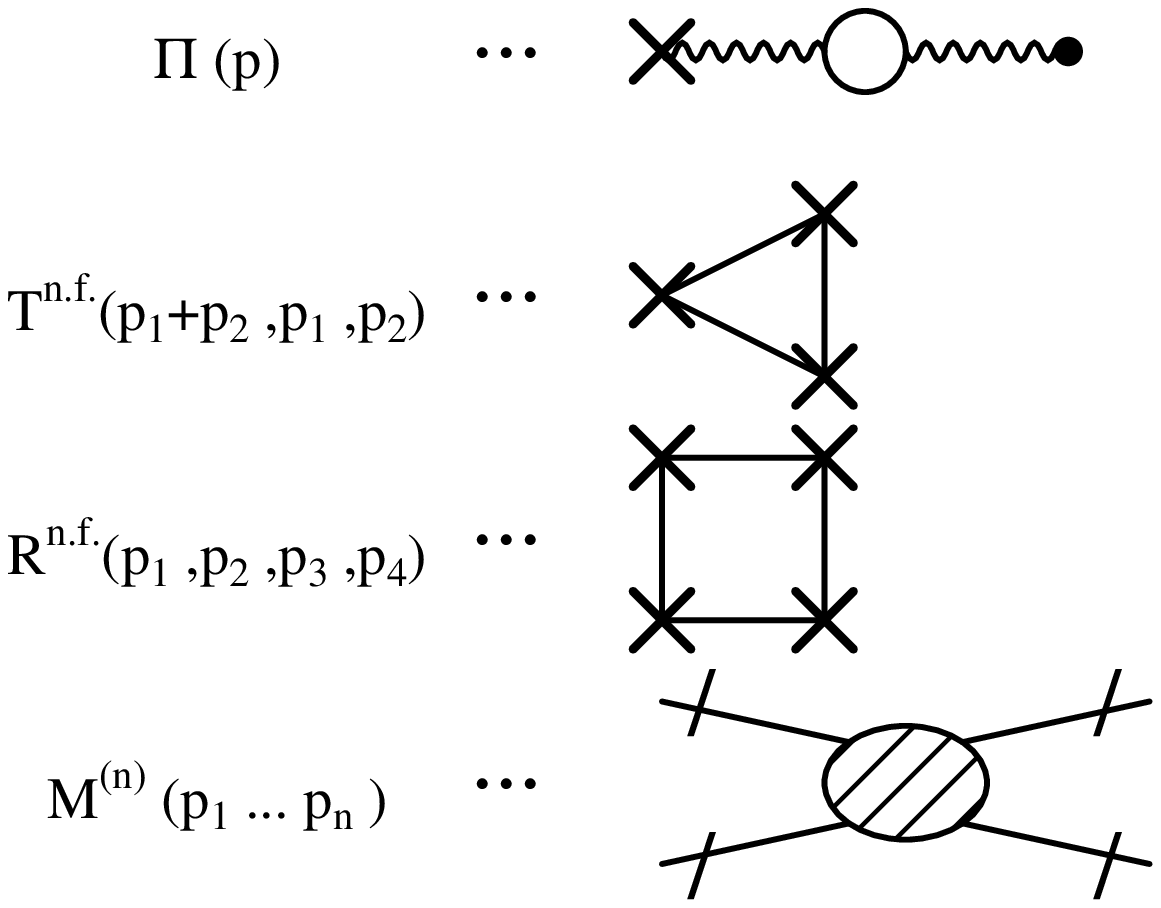}}{\fillinggrid
\at(5.9\pscm;-0.5\pscm){Fig. 4}}$$ 

\vspace{0.1cm}

where $M^{(n)}$, of course, should have $n$ external (amputated) boson lines.

For the three-point function e.g. we find the Dyson-Schwinger equation
\bdi
M^{(3)}(p_1 +p_2 ,p_1 ,p_2 )=(2\sqrt{\pi})^3 [m\langle P\rangle_m +m^2
\wt{\langle SP\rangle}_m^c (p_1 +p_2 )
\edi
\beq
+m^2 \wt{\langle SP\rangle}_m^c (p_1)+m^2 \wt{\langle SP\rangle}_m^c (p_2) + 
m^3 \wt{\langle PPP\rangle}_m^c (p_1 +p_2 ,p_1 ,p_2) ]
\eeq
where $m\langle P\rangle_m$ and $m^2 \wt{\langle SP\rangle}_m^c (p)$ are
analogous to (24), (25) whereas the last term is given by
\beq
m^3 \wt{\langle PPP\rangle}_m^c (p_1 +p_2 ,p_1 ,p_2)=P_iP_jP_k {\cal G}_{ii'}
{\cal G}_{jj'}{\cal G}_{kk'}T_{i' j' k'}(p_1 +p_2 ,p_1 ,p_2 )
\eeq
and $T_{ijk}$ is given by the graphs of Fig. 5

$$\psannotate{\psboxscaled{650}{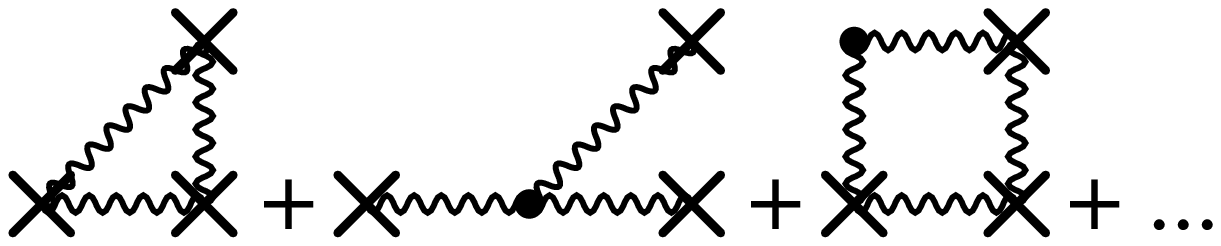}}{\fillinggrid
\at(5.9\pscm;0.0\pscm){Fig. 5}}$$ 

\vspace{0.1cm}

The essential point is that $M^{(3)}$, again, may be reexpressed entirely in
terms of non-factorizable $n$-point functions, namely
\bdi
M^{(3)}(p_1 +p_2 ,p_1 ,p_2)=(2\sqrt{\pi})^3 P_iP_jP_k \Pi_{ii'}(p_1 +p_2 )
\Pi_{jj'}(p_1 )\Pi_{kk'}(p_2) \cdot
\edi
\beq
\cdot \Bigl( {\cal G}_{i' j' k'}+ {\cal G}_{i' l}
{\cal G}_{j' m}{\cal G}_{k' n}T^{\rm n.f.}_{lmn}(p_1 +p_2 ,p_1 ,p_2)\Bigr)
\eeq
or, graphically (see Fig. 6)

$$\psannotate{\psboxscaled{500}{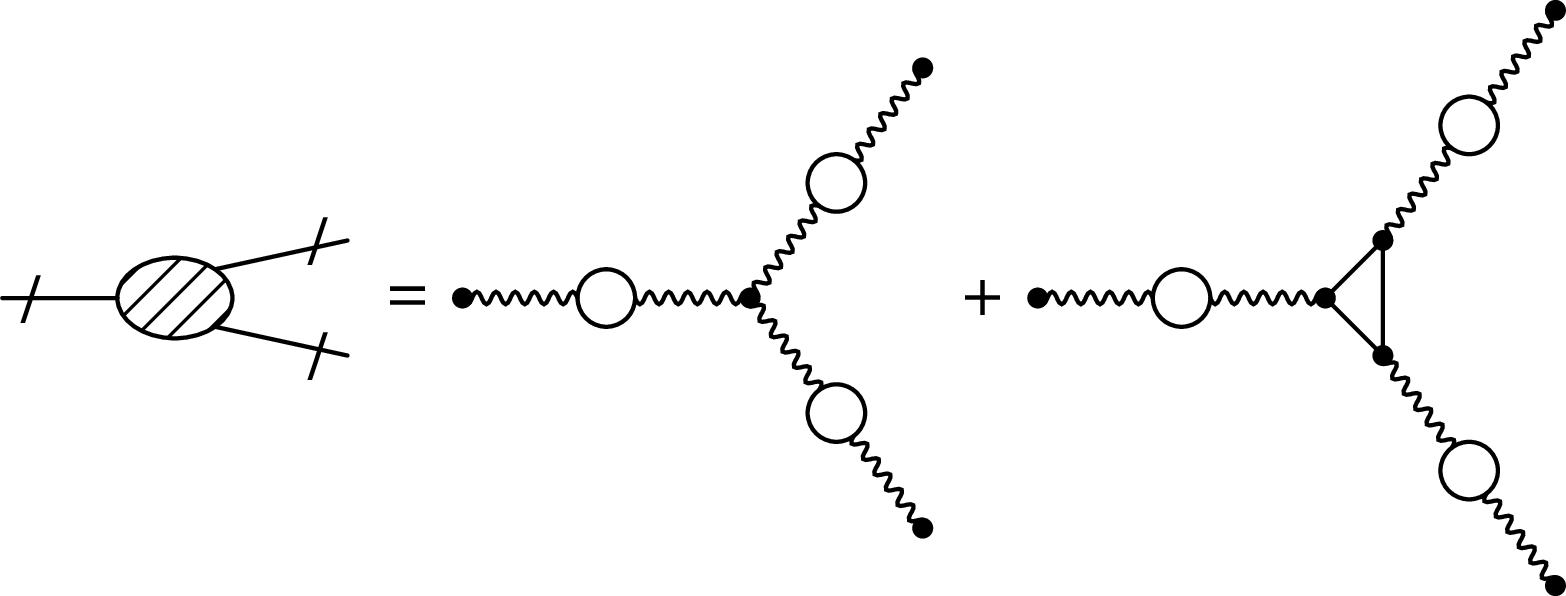}}{\fillinggrid
\at(10.9\pscm;-0.5\pscm){Fig. 6}}$$ 

where the non-factorizable three-point function $T_{\rm n.f.}$ is given by
Fig. 7

$$\psannotate{\psboxscaled{700}{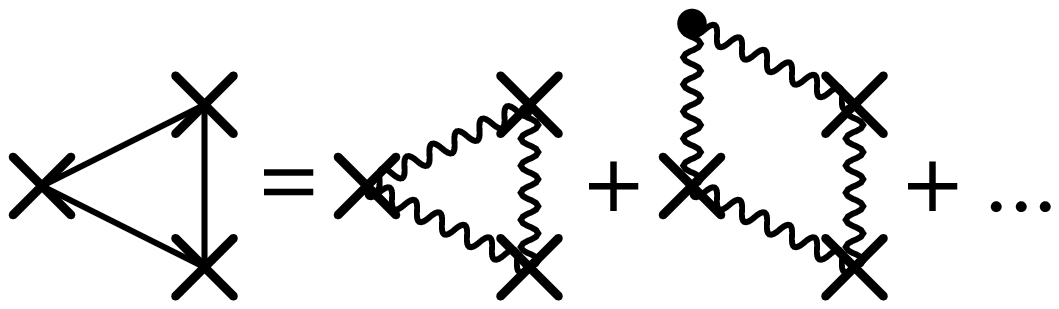}}{\fillinggrid
\at(5.9\pscm;-0.5\pscm){Fig. 7}}$$ 

\vspace{0.1cm}

The actual validity of (28), Fig. 6 has to be checked by a closer inspection of
the Feynman graphs (it is just tedious combinatorics).

We find that the non-factorizable $n$-point functions in our theory play a role
analogous to the 1PI Green functions in other theories.

The four-point function $M^{(4)}$ may be treated along similar lines. Again,
the Dyson-Schwinger equation allows to express $M^{(4)}$ in terms of $\langle
P\ldots \rangle_m^c$ and $\langle S\ldots \rangle_m^c$ 
$n$-point functions, which we do
not display (the explicit expression is written down in \cite{GBOUND}).
Further, $M^{(4)}$ may be reexpressed in terms of non-factorizable $n$-point
functions and reads
\bdi
M^{(4)}(p_1 ,\ldots ,p_4)=(4\pi )^2 P_iP_jP_kP_l
\Pi_{ii'}(p_1)\Pi_{jj'}(p_2) \Pi_{kk'}(p_3) \Pi_{ll'}(p_4)\Bigl[ {\cal
G}_{i' j' k' l'}
\edi
\bdi 
+ {\cal G}_{i' m}{\cal G}_{j' m'}{\cal G}_{k' n}{\cal G}_{l' n'} R^{\rm
n.f.}_{mm' nn'}(p_1 ,p_2 ,p_3 ,p_4)
\edi
\bdi
+\Bigl( {\cal G}_{i' j' m}(\Pi_{mm'}(p_1 +p_2 )-\delta_{mm'})\delta_{m'
k' l'} +\mbox{ perm. }\Bigr)
\edi
\bdi
+\Bigl( {\cal G}_{i' j' m}\Pi_{mm'}(p_1 +p_2)T^{\rm n.f.}_{m' nn'}
(p_1 +p_2 ,p_3,
p_4){\cal G}_{nk'}{\cal G}_{n' l'} + \mbox{ perm. }\Bigr)
\edi
\beq
+\Bigl( {\cal G}_{i' m}{\cal G}_{j' m'}T^{\rm n.f.}_{mm' n}(p_1 +p_2 ,p_1
,p_2){\cal G}_{nn'}\Pi_{n' r}(p_1 +p_2)T^{\rm n.f.}_{rr' s}(p_3 +p_4 ,p_3
,p_4){\cal G}_{r' k'}{\cal G}_{sl'} + \mbox{ perm. }\Bigr) \Bigr]
\eeq
where momentum conservation requires $p_1 +p_2 =p_3 +p_4$. Graphically, this
identity may be depicted like in Fig. 8.

$$\psannotate{\psboxscaled{500}{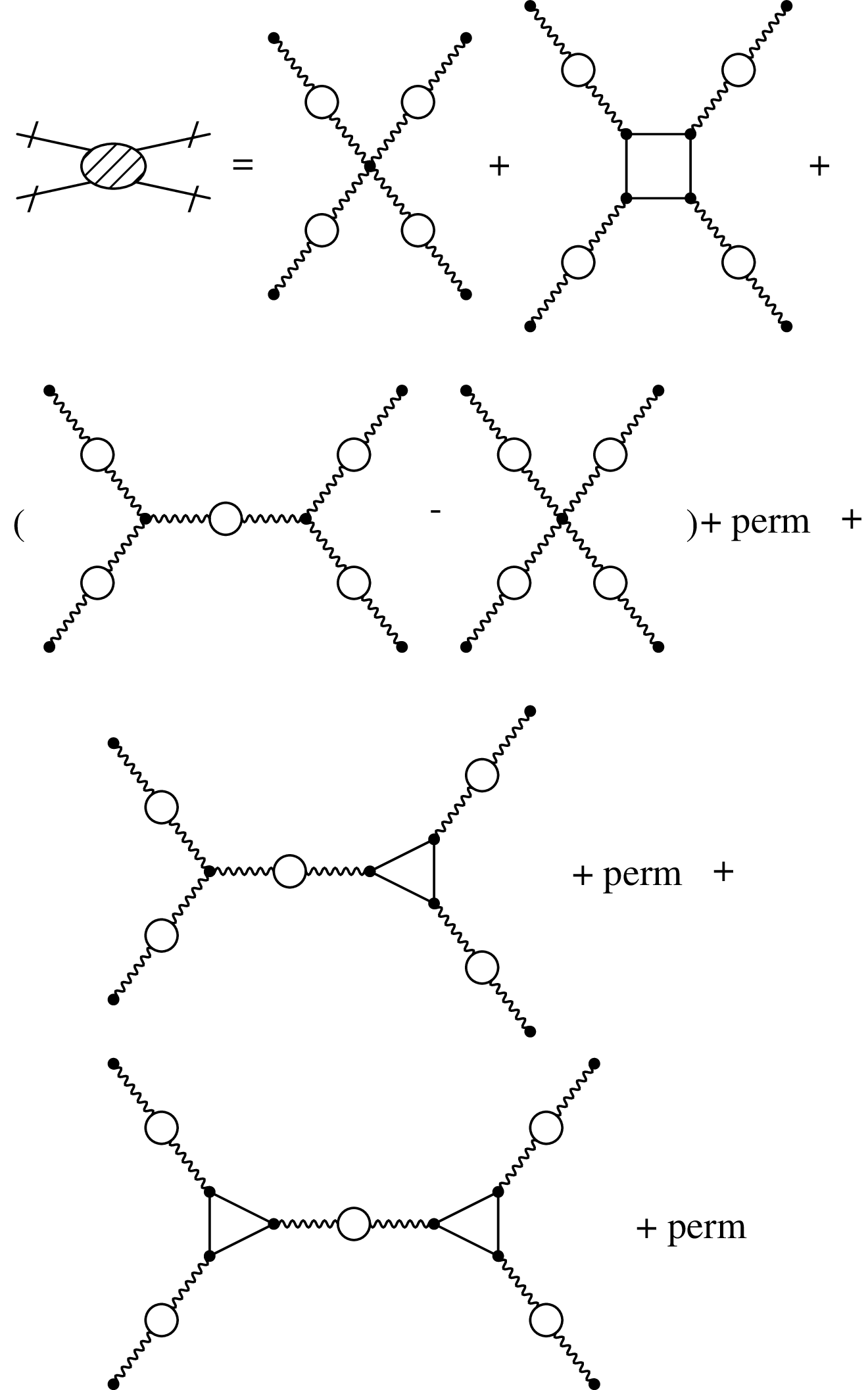}}{\fillinggrid
\at(10.9\pscm;-0.5\pscm){Fig. 8}}$$ 

The permutations in Fig. 8 contain all attachments of the external
$\Pi (p_i)$ lines that are topologically distinct (i.e. 3, 6 and 3
permutations).

Observe that in each of the third type diagrams of Fig. 8 
the lowest order diagram has to
be subtracted in order to avoid an overcounting (this is so because $\Pi (p)$
contains the lowest order, ${\cal G}\Pi (p)={\cal G} +o(g_\theta^2)$).

\section{Bound-state structure}

Before turning to the actual scattering processes we should shortly discuss
some other physical properties of the model. Actually quite a lot of physical
information may be obtained from the two-point function $\Pi (p)$, (19), Fig.
2b. First, observe that the $\Pi (p)$ propagator also occurs in higher bosonic
$n$-point functions. E.g. for $M^{(4)}$ (Fig. 8), when one takes the third
type of diagrams and inserts the lowest order ($\Pi \sim {\bf 1}$) for the four
{\em external} $\Pi (p_i)$ lines, there remains precisely an internal $\Pi (p_1
+p_2)$ propagator (times ${\cal G}$). Therefore it is not a surprize that we
can provide information on higher bosonic states, too, from $\Pi (p)$. In fact,
most of the information may be inferred from the denominator $N(p)$, (20), of
$\Pi (p)$. The zeros of the real part of $N(p)$ will give all the bound-state
masses of the theory -- at least in leading order -- and the imaginary parts
will give the corresponding decay widths, as was discussed in 
\cite{GBOUND,DECAY,THREE}. We give a short review of these results, because we
need them in the sequel. The denominator $N(p)$ depends on two functions
$\alpha$, $\beta$, which are in lowest order
\beq
\alpha (g_\theta ,p)=g_\theta \wt E_+ (p) \quad ,\quad \beta (g_\theta ,p)=
g_\theta \wt E_- (p) \quad ,\quad g_\theta =m\frac{\Sigma}{2}e^{i\theta}
\eeq
and the $\wt E_\pm (p)$ are the exponentials of bosonic propagators,
\beq
\wt E_\pm (p)=\sum_{n=1}^\infty (\pm 1)^n d_n (p) \quad ,\quad d_n (p):=
\frac{(4\pi)^n}{n!}\wt{D^n_{\mu_0}}(p) .
\eeq
The higher order terms which we ignored in (30) have some important effects on
the lowest order expression (31) that may not be ignored. First, they cause
corrections on the individual boson lines in (31) that shift all the internal
bosons from the bare Schwinger mass $\mu_0$ to the physical Schwinger mass
$\mu$ (to be displayed in the next section; 
actually this shift does not occur for the
one-boson part of $\wt E_\pm (p)$, $d_1 (p)$, because mass corrections to $d_1
(p)$ are given by factorizable graphs that are excluded from $\alpha$, $\beta$,
see \cite{GBOUND}). Therefore we redefine the $d_n (p)$ ($n\ge 2$) for the rest
of the paper to be
\beq
d_n (p):=\frac{(4\pi)^n}{n!}\wt{D^n_\mu}(p) .
\eeq
The $d_n (p)$ are just $n$-boson blobs (see Fig. 9 for $d_2$, $d_3$)

$$\psannotate{\psboxscaled{700}{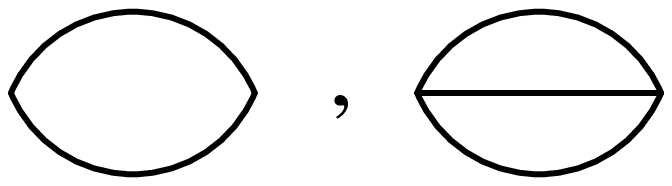}}{\fillinggrid
\at(2.9\pscm;-0.5\pscm){Fig. 9}}$$ 

\vspace{0.1cm}

and have the following properties: at $s=-p^2 =(n\mu)^2$, $d_n (p)$ has a
singularity (real particle production threshold), and above this threshold it
has an imaginary part. Therefore, slightly below the threshold $(n\mu)^2$, $d_n
(p)$ is large enough to balance the coupling constant and make the real part of
$N(p)$, (20), vanish,
\beq
m\Sigma \cos\theta d_n (p)\sim 1+o(m)
\eeq
and, therefore, causes an $n$-boson bound state. At the position of the
two-boson bound state, $s=M_2^2 =4\mu^2 -\Delta_2$, $N(p)$ has no imaginary
part and, therefore, the two-boson bound state is stable. At the three-boson
bound-state mass $M_3$, $d_2 (s=M_3^2)$ has an imaginary part and, therefore, a
decay into two Schwinger bosons (with mass $\mu$) is possible. For higher
$n$-boson bound states the functions $d_2
,\ldots ,d_{n-1}$ have imaginary parts at $M_n^2$,
therefore decays into $2,\ldots ,n-1$ Schwinger bosons are
possible (\cite{GBOUND,DECAY,THREE}).

However, this can not yet be the whole story. To understand why, look at the
lowest order contribution to the three-point function, Fig. 6, with one
incoming $\Pi (p_1)$, one vertex and two outgoing $\Pi (p_2)$, $\Pi (p_3)$.
Suppose the incoming $\Pi (p_1)$ is at the mass $-p_1^2 =M_n^2$ of a
sufficiently heavy unstable bound state. For a decay into stable final
particles all the stable mass poles of $\Pi (p_2)$, $\Pi (p_3)$ are possible.
But by our above arguments the mass pole of the stable $M_2$ particle is
present in $\Pi (p_i)$ as well as the mass pole of the Schwinger boson $\mu$.
Therefore, Fig. 6 describes decays into $M_2$ particles as well as $\mu$
particles. On the other hand, we did not find imaginary parts (up to now) in
$N(p)$ that describe decays into some $M_2$, so obviously something is missing.

The $M_2$ bound state itself was found by a resummation, therefore it is a
reasonable idea to use the higher order contributions to $\alpha$, $\beta$ for
a further resummation. $\alpha$ and $\beta$ are just components of the
non-factorizable propagator ${\cal A}(p)$, (16), so let us investigate it more
closely.

By a partial resummation we may find the following contribution to ${\cal
A}(p)$,
\beq
H_{ii'}(p):=\int\frac{d^2 q}{(2\pi)^2} \delta_{ijk}{\cal A}_{jj'}(q){\cal
G}_{j' l}\Pi_{ll'}(q){\cal A}_{l' k'}(q){\cal A}_{km}(q-p) \delta_{i' k'
m}.
\eeq
This is just a blob where ${\cal A}(q-p)$ runs along one line, the other terms
run along the other line. We want to discuss the $\mu$-$M_2$ contribution,
therefore we substitute ${\cal A}(q-p)$ by its lowest order, one-boson part,
\beq
{\cal A}(q-p)\sim 4\pi \wt D_\mu (q-p)
\left( \begin{array}{cc} 1 & -1 \\ -1
& 1 \end{array} \right) 
\eeq
and the two further ${\cal A}(q)$ by their lowest order contribution
\beq
 {\cal A}(q)\sim 
\left( \begin{array}{cc} \wt E_+ (q) & \wt E_- (q) \\ \wt E_- (q)
& \wt E_+ (q) \end{array} \right) 
\eeq
The resummation that we need in (34) is taken into account by $\Pi (q)$. 
With these restrictions $H(p)$ corresponds to the graph of Fig. 10

$$\psannotate{\psboxscaled{700}{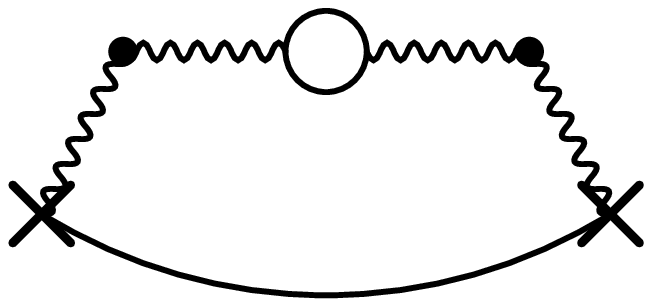}}{\fillinggrid
\at(2.9\pscm;-0.7\pscm){Fig. 10}}$$ 

\vspace{0.5cm}

Observe that all internal bosons may be renormalized to their physical masses
$\mu$, because this does not spoil non-factorizability in Fig. 10. The two
factors ${\cal A}(q)$ in (34) are necessary in order to avoid an overcounting,
but they cannot influence the presence of higher poles in Fig. 10.

Now suppose that $H(p)$ is at the $M_2 +\mu$-threshold, $s=-p^2=(M_2 +\mu)^2$.
Then $\wt D_\mu (q-p)$ is at its $\mu$-singularity and $\Pi (q)$ at its
$M_2$-singularity, and Fig. 10 corresponds (up to a normalization) to a
$\mu$-$M_2$ two-boson loop, i.e. Fig. 10 may effectively be substituted by 
Fig. 11,

$$\psannotate{\psboxscaled{750}{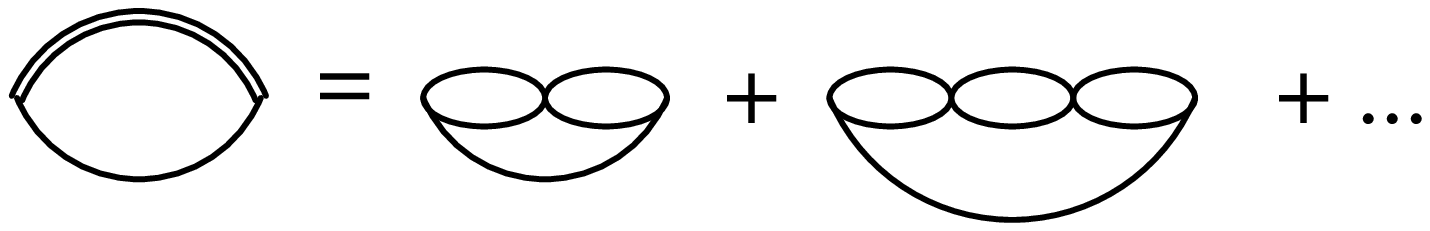}}{\fillinggrid
\at(6.9\pscm;-0.5\pscm){Fig. 11}}$$ 

\vspace{0.3cm}

where the double line represents the two-boson bound-state propagator.

Therefore, $H(p)$ is singular at $-p^2=(M_2 +\mu)^2$, and has a large real part
slightly below and a large imaginary part slightly above this threshold. As a
consequence, when the contribution of $H(p)$ to $\alpha (p)$ is
taken into account in the denominator $N(p)$, (20), it will give rise to a
further $\mu $-$M_2$ bound state slightly below $s=(M_2 +\mu)^2$. Further it
will open the $\mu$-$M_2$ decay channel at $s=(M_2 +\mu)^2$.

Now suppose we put $\Pi (q)$ in (34) on a higher (unstable) bound-state mass
$M_n$, $n>2$.
Then in the denominator $N(q)$ of $\Pi (q)$ (see (20)) the real part again
vanishes, but there remains an imaginary part. Therefore, $H(p)$ is finite and
imaginary at $s=-p^2 =(M_n +\mu)^2$ and {\em cannot} give rise to a $\mu$-$M_n$
bound-state formation.

Further, because there is no threshold singularity at $s=(M_n +\mu)^2$, this
means that {\em no} new decay channel opens at that point (i.e. the imaginary
part of $H(p)$ varies smoothly around $s\sim (M_n +\mu^2$)), which simply means
that the unstable higher $n$-boson bound states are no possible final states
(of course, they are possible as intermediate resonances).

We could substitute the one-boson line in Fig. 10 by another ${\cal A}{\cal
G}\Pi {\cal A}$ line and would find that this graph behaves like a $M_2$-$M_2$
blob near $s=(M_2 +M_2)^2$, and we could allow for even more ${\cal A}{\cal G}
\Pi {\cal A}$ lines. The physical picture that evolves from these
considerations is like follows: in addition to the unstable $n$-boson bound
states there exist further (unstable) bound states that are composed of
Schwinger bosons $\mu$ and (stable) two-boson bound states $M_2$. 
Further, the
unstable bound states may decay into all combinations of $\mu$ and $M_2$
particles that are possible kinematically. The imaginary parts of the
corresponding $n$-particle blobs (where particle means $\mu$ or $M_2$) are
large near their thresholds, therefore there is a kinematical tendency to rise
the decay probabilities for decays with {\em small} kinetic energy. 
This is not so
surprizing, because in $1+1$ dimensions the phase space "volume" does not grow
with kinetic energy.

\section{Bound-state masses and decay widths}

Up to now we discussed the physical properties of the model on a qualitative
level, but for the further discussion we need some explicit results, too.

The mass pole equation in lowest order reads
\beq
1=(g_\theta +g_\theta^* ){\rm Re} \wt E_+ (p)
\eeq
or, for the $n$-boson bound state
\beq
f_n (p):=1-m\Sigma\cos\theta d_n (p) =0
\eeq
and has the three lowest solutions ($n=1, 2,3$; we only display the leading
order corrections)
\beq
M_1^2 \equiv \mu^2 =\mu_0^2 +\Delta_1   \quad ,\quad \Delta_1 =4\pi
m\Sigma\cos\theta
\eeq
\beq
M_2^2 =4\mu^2 -\Delta_2 \quad ,\quad \Delta_2 =\frac{4\pi^4 m^2 \Sigma^2 \cos^2
\theta}{\mu^2}
\eeq
\beq
M_3^2 =9\mu^2 -\Delta_3 \quad ,\quad \Delta_3 \simeq 
6.993 \mu^2 \exp (-0.263\frac{\mu^2}{m\Sigma \cos\theta})
\eeq
(these masses have already been computed in \cite{GBOUND};
there is, however, a numerical error in the $M_2$ mass formula in
\cite{GBOUND}). The $f_n (p)$ of (38) may be expanded in Taylor series about
their respective mass poles $s-M_n^2$. The leading coefficients ($s=-p^2$)
\beq
f_n (p)\simeq c_n (s-M_n^2)
\eeq
which we need for the residues of the mass poles, are (see \cite{THREE} for a
computation)
\beq
c_1 =\frac{1}{4\pi m\Sigma\cos\theta}=\frac{1}{\Delta_1}
\eeq
\beq
c_2 =\frac{\mu^2}{8\pi^4 (m\Sigma\cos\theta)^2}=\frac{1}{2\Delta_2}
\eeq
\beq
c_3 =\frac{m\Sigma\cos\theta}{0.263 \mu^2 \Delta_3}
\eeq
The next things we need are the residues of the propagator $\Pi (p)$ at the
two lowest mass poles.
 Inserting the lowest order ($\alpha \sim g_\theta \wt
E_+ $ etc.) into formula (19) and using the pole equation (38) we find at once
(here $n=1,2$)
\beq
\Pi (s\sim M_n^2 ) \sim \frac{1}{(g_\theta +g^*_\theta )c_n
(s-M_n^2 )} \left( \begin{array}{cc} g_\theta & (-1)^n g_\theta^* 
\\ (-1)^n g_\theta & g_\theta^* \end{array} \right) 
\eeq
For the computation of the $\mu$-$M_2$ bound state we need, in addition, the
matrix ${\cal A}$, (16), at the $n$-boson mass poles (here $n\ne 1$, because
there ${\cal A}$ itself has a pole (46)),
\beq
{\cal A}(s=M_n^2 )=\frac{1}{g_\theta +g^*_\theta } \left( \begin{array}{cc} 1 &
(-1)^n \\ (-1)^n & 1 \end{array} \right) 
\eeq
Using these results we find for the matrix $H(p)$, (34), near the $\mu$-$M_2$
threshold
\bdi
H_{ii'}(-p^2 \sim (M_2 +\mu)^2)\sim \int\frac{d^2 q}{(2\pi)^2} \delta_{ijk}
\frac{1}{g_\theta +g^*_\theta } \left( \begin{array}{cc} 1 & 1 \\ 1 & 1
\end{array} \right)_{jj'} \left( \begin{array}{cc} g_\theta & 0 \\ 0 &
g^*_\theta \end{array} \right)_{j' l} \frac{1}{(g_\theta +g^*_\theta)c_2 (-q^2
-M_2^2 )} \cdot
\edi
\bdi
\cdot \left( \begin{array}{cc} g_\theta & g_\theta^* \\ 
g_\theta & g_\theta^* \end{array} \right)_{ll'}
\frac{1}{g_\theta +g^*_\theta } \left( \begin{array}{cc} 1 & 1 \\ 1 & 1
\end{array} \right)_{l' k'} \frac{1}{(g_\theta +g^*_\theta )c_1 (-(q-p)^2
-\mu^2 )} \left( \begin{array}{cc} 1 & -1 \\ -1 & 1 \end{array} \right)_{km}
\delta_{i' k' m} =
\edi
\beq
\delta_{ijk} \left( \begin{array}{cc} 1 & 1 \\ 1 & 1 \end{array} \right)_{jj'}
\left( \begin{array}{cc} 1 & -1 \\ -1 & 1 \end{array} \right)_{kk'} \delta_{i'
j' k'} \int\frac{d^2 q}{(2\pi)^2} \frac{1}{(g_\theta +g^*_\theta )^2 c_1 c_2
(-q^2 -M_2^2 )(-(q-p)^2 -\mu^2 )}
\eeq
The contribution of $H_{ii'}$ to $\alpha (g_\theta) +\alpha (g^*_\theta )$ in
the denominator $N(p)$, (20), is
\bdi
g_\theta H_{++} (p) +g^*_\theta H_{--} (p) =: (g_\theta +g^*_\theta )d_{1,1}
(p)=
\edi
\bdi
(g_\theta +g^*_\theta )\int\frac{d^2 q}{(2\pi)^2}\frac{8\pi^4
m\Sigma\cos\theta}{\mu^2 (q^2 +M_2^2 )} \frac{4\pi}{(p-q)^2 +\mu^2 }=
\edi
\bdi
 \frac{32\pi^5 m^2 \Sigma^2 \cos^2 \theta}{2\pi \mu^2\bar w(s ,M_2^2
,\mu^2 )}\Bigl( \pi +
\edi
\beq
\arctan \frac{2s}{\bar w(s ,M_2^2 ,\mu^2 ) 
-\frac{1}{\bar w(s ,M_2^2 ,\mu^2 )}(s +\mu^2 -M_2^2)(s -\mu^2 +M_2^2 )}
\Bigr)
\eeq
\beq
\bar w(x,y,z):=(-x^2 -y^2 -z^2 +2xy+2xz+2yz)^{\frac{1}{2}}
\eeq
where $s=-p^2 \sim (\mu +M_2)^2 $. The $\mu$-$M_2$ bound-state mass fulfills
the equation
\beq
1=(g_\theta +g^*_\theta )d_{1,1}(p)
\eeq
with the solution in leading order (here $M_{1,1}$ denotes the $\mu$-$M_2$
bound-state mass)
 \beq
M_{1,1}^2 =(\mu +M_2)^2 -\Delta_{1,1}\quad ,\quad \Delta_{1,1}=
\frac{32\pi^{10}(m\Sigma\cos\theta)^4}{\mu^6}
\eeq
which is valid for sufficiently small $\Delta_{1,1}$. $M_{1,1}$ was computed
from a two-boson blob (Fig. 11), like $M_2$, 
therefore the Taylor coefficient of
$(s-M_{1,1}^2 )$ is analogous to $c_2$, equ. (44),
\beq
c_{1,1} =\frac{1}{2\Delta_{1,1}}=\frac{\mu^6}{64\pi^{10}(m\Sigma\cos\theta)^4}.
\eeq
Further, the above equ.
(48) shows that the $\mu$-$M_2$ blob $d_{1,1}(p)$ enters into the functions
$\alpha$, $\beta$ of ${\cal A}$, (16), like any other odd $n$-boson blob $d_n
(p)$ ($d_{1,1}(p)$ consists of three Schwinger bosons and is, therefore, parity
odd).

In the sequel we will be interested in the behaviour of the exact two-point
function $\Pi (p)$, equ. (19), and its denominator $N(p)$ 
in the vicinity of the
mass poles $s\sim M_n^2$. Using the first order approximation (30) 
for $\alpha$,
$\beta$ we find for $N(p)$ ( e.g. in the vicinity of $s\sim M_3^2$ for
definiteness)
\beqa
N(p) & \simeq & 1-m\Sigma \cos\theta\wt E_+ (p) +\frac{m^2 \Sigma^2}{4} (\wt
E_+^2 (p) -\wt E_-^2 (p)) \no \\
& = & 1-m\Sigma\cos\theta (d_1 (p) +d_2 (p) +d_{1,1} (p) +d_3 (p) +\ldots )+ 
\no \\
&& m^2 \Sigma^2 \Bigl( d_1 (p)(d_2 (p) +d_4 (p) +\ldots ) +d_{1,1}(p) (d_2 (p)
+ d_4 (p) +\ldots ) + \no \\ 
&& d_3 (p) (d_2 (p) +d_4 (p) +\ldots )+\ldots \Bigr)
\eeqa
where we included the $\mu$-$M_2$ blob $d_{1,1}$, as discussed above, because
we need it for the subsequent discussion (we ignore, for the moment, 
higher $M_2$
blobs that are, in principle, present). Near $s=M_3^2$ the real part of (54) is
given by $c_3 (s-M_3^2 )$ and we find
\beqa
N(p) & \sim & c_3 (s-M_3^2 ) -im\Sigma\cos\theta ({\rm Im} d_2 (s\sim M_3^2 )
+{\rm Im} d_{1,1}(s\sim M_3^2 )) + \no \\
&& im^2 \Sigma^2 d_3 (s\sim M_3^2 ){\rm Im} d_2
(s\sim M_3^2 ) +o(m^2) \no \\
&=& c_3 (s-M_3^2 )-im\Sigma (\cos\theta -\frac{1}{\cos\theta}){\rm Im} d_2
(M_3^2 ) -im\Sigma\cos\theta \, {\rm Im} d_{1,1}(M_3^2 ) +o(m^2) \no \\
&&
\eeqa
where we used $d_3 (M_3^2)\sim \frac{1}{m\Sigma\cos\theta}$, see (38).

This computation may be generalized and tells us that parity forbidden
imaginary parts (decay channels) acquire a factor $(\cos\theta
-\frac{1}{\cos\theta})$, whereas parity allowed imaginary parts have the usual
$\cos\theta$ factor.

Remark:
There seems to be something wrong with the sign of the parity forbidden 
imaginary part (the $d_2$ term). Actually the sign is o.k.
and the problem is a remnant of the Euclidean conventions that are implicit in
the whole computation (see e.g. \cite{ABH,MSSM}). In these conventions $\theta$
is imaginary and therefore $(\cos\theta -\frac{1}{\cos\theta}) \ge 0$. 
Of course,
this is not a reasonable convention for a final result. When performing the
whole computation in Minkowski space and for real $\theta$, roughly speaking,
the roles of $E_+$ and $E_-$ are exchanged in (54). This gives an additional
relative sign between parity even and odd $n$-boson propagators and, therefore,
changes the factor of $d_2$ to $(\frac{1}{\cos\theta} -\cos\theta )$, which is
$\ge 0$ for real $\theta$.
We will keep this remark in mind and express the final results in Minkowski
space and for real $\theta$. 

From the imaginary parts of (55) 
one is able to compute the partial decay widths
of the three-boson bound state, and in a similar way one may compute the decay
widths of the other unstable bound states (see \cite{DECAY,THREE}). The
explicit results for the $M_3$ and $M_{1,1}$ decay widths read (in Minkowski
space and for real $\theta$)
\beq
\Gamma_{M_{1,1}\ra 2\mu}=
\frac{2^8 \pi^{12} (m\Sigma\cos\theta)^5}{9\sqrt{5}\mu^9}
(\frac{1}{\cos^2 \theta} -1) \simeq 21340 \mu(\frac{m\cos\theta}{\mu})^5
(\frac{1}{\cos^2 \theta} -1)
\eeq
\beq
\Gamma_{M_3 \ra 2\mu}=0.263 \frac{4\pi^2
\Delta_3}{9\sqrt{5}\mu}(\frac{1}{\cos^2 \theta}-1)
\simeq 3.608 \mu (\frac{1}{\cos^2 \theta}-1)
\exp (-0.929\frac{\mu}{m\cos\theta})
\eeq
\beq
\Gamma_{M_3 \ra M_2 +\mu}=0.263\frac{4\pi^3 \Delta_3}{3\sqrt{3}\mu}
\simeq 43.9 \mu \exp (-0.929\frac{\mu}{m\cos\theta})
\eeq
We will need expression (55) for our scattering discussion.

\section{Two-dimensional kinematics}

Next we need some basic facts about two-dimensional kinematics. We will
restrict our discussion to elastic scattering. Suppose we have two incoming
particles with masses $M_1$, $M_2$ and momenta $p_1$, $p_2$, and two outgoing
particles, again with masses $M_1$,  $M_2$, and with momenta $p_3$, $p_4$.
Momentum conservation requires 
\beq
p:=p_1 +p_2 =p_3 +p_4
\eeq
and all momenta are {\em Minkowskian} in the sequel. In the center of mass
system we may write 
\bdi
p_1 ={\sqrt{k^2 +M_1^2} \choose k} \quad ,\quad p_2 ={\sqrt{k^2 +M_2^2} \choose
-k}
\edi
\beq
p_3 ={\sqrt{k^2 +M_1^2} \choose \pm k} \quad ,\quad p_4 ={\sqrt{k^2 +M_2^2} 
\choose \mp k}
\eeq
where in $p_3$, $p_4$ the first sign is for transmission, the second sign is
for reflexion. For the kinematical variables we find for transmission
\beqa
s=(p_1 +p_2 )^2 &=& 2k^2 +M_1^2 +M_2^2 +2\sqrt{(k^2 +M_1^2)(k^2 +M_2^2)} \no \\
t_T =(p_1 -p_4 )^2 &=& -2k^2 +M_1^2 +M_2^2 -
2\sqrt{(k^2 +M_1^2)(k^2 +M_2^2)} \no \\
u_T =(p_1 -p_3)^2 &=&0
\eeqa
and for reflexion
\beqa
s=(p_1 +p_2 )^2 &=& 2k^2 +M_1^2 +M_2^2 +2\sqrt{(k^2 +M_1^2)(k^2 +M_2^2)} \no \\
t_R =(p_1 -p_4 )^2 &=& 2k^2 +M_1^2 +M_2^2 -
2\sqrt{(k^2 +M_1^2)(k^2 +M_2^2)} \no \\
u_R =(p_1 -p_3)^2 &=& -4k^2
\eeqa
When the two masses are equal, the two particles are identical in our theory
and the discrimination between transmission and reflexion does not make sense.
The kinematical variables turn into
\beqa
s=(p_1 +p_2 )^2 &=& 4(k^2 +M^2) \no \\
t =(p_1 -p_4 )^2 &=& -4k^2 \no \\
u =(p_1 -p_3)^2 &=& 0
\eeqa
The elastic scattering cross section of two particles is given by
\beq
\sigma_{M_a M_b \ra M_a M_b}(s)=\frac{C_{\rm sym}|{\cal M}(s)|^2}{2w^2 (s,M_a^2
,M_b^2)}
\eeq
\beq
w(x,y,z)=(x^2 +y^2 +z^2 -2xy-2xz-2yz)^{\frac{1}{2}}
\eeq
where ${\cal M}$ is the transition matrix element and $C_{\rm sym}$ is a
symmetry factor that takes into account identical particles in the final state
($C_{\rm sym} =\frac{1}{n_1 !n_2 !}$ for $n_1$ particles $M_1$ and $n_2$
particles $M_2$ in the final state). As it stands, expression (64) holds
provided that the initial and final particle propagators are normalized in the
usual fashion ($\sim \frac{1}{s-M_i^2}$). Otherwise, (64) is multiplied by the
normalization factors (the residues of the propagators).

Unitarity relates the scattering cross section to the imaginary parts of some
graphs, therefore let us write down the imaginary part of the two-boson blob
for later convenience,
\beqa
{\rm Im \,}\wt{(D_{M_1}D_{M_2})}(s)&=& {\rm Im \,}\int \frac{d^2 q}{(2\pi)^2}
\frac{-1}{q^2 +M_1^2}\frac{-1}{(p-q)^2 +M_2^2} \no \\
&=& \frac{1}{2w(s,M_1^2 ,M_2^2 )}
\eeqa

\section{Scattering processes}

Finally we are prepared for a discussion of scattering. Let us focus for the
moment on the lowest order graph of Fig. 8 for the four-point function (29). 
It consists of
four external exact propagators $\Pi (p_i)$ and a simple vertex as the lowest
order transition matrix element. The $\Pi (p_i)$ contain two stable-particle
mass poles, $\mu$ and $M_2$, therefore this graph describes $\mu$ and $M_2$
scattering (this remains true for higher order contributions; as a consequence,
the same transition matrix elements contribute to $\mu$ and $M_2$ scattering
processes, and they may only differ by some kinematical and normalization
factors).

Let us consider elastic scattering of two Schwinger bosons for definiteness.
Then each external $\Pi (p_i)$ propagator is odd and contributes to the graph
like ($s_i =-p_i^2$) 
\beq
\Pi_{jk}(s_i =\mu^2)P_k =
\frac{4\pi (g_\theta +g^*_\theta)}{s_i -\mu^2}{1
\choose -1}_j
\eeq
Here we face the problem that the first graph of Fig. 8 is already of fifth 
order, because each propagator $\Pi (s_i)$ has an external vertex. We just
omit these external vertices (i.e. we omit the factor $(g_\theta +g^*_\theta)$
in (67) for each propagator), because we want to discuss first order 
scattering. Doing so, we find for this graph 
\beq
P_{j_1}P_{j_2}P_{j_3}P_{j_4}
\delta_{j_1 j_2 k_1}{\cal G}_{k_1 k_2}\delta_{j_3 j_4 k_2}
\prod_{i=1}^4 \frac{4\pi}{s_i -\mu^2}
= (g_\theta +g^*_\theta )\prod_{i=1}^4 \frac{4\pi}{s_i -\mu^2}
\eeq
i.e. each $\mu$ propagator has a residue $4\pi$. In order to obtain the
transition matrix element one has to amputate the external boson propagators in
the usual LSZ fashion. When the propagators are normalized by $r_1 =4\pi$, the
bosons themselves are normalized by $\sqrt{4\pi}$, which has to be divided out
for each amputation. This leaves a factor $\sqrt{4\pi}$ for each external boson
in the transition matrix element. However, the squared transition matrix element
enters the scattering cross section, therefore the net effect on the cross
section is a multiplication by the corresponding propagator residue $r_i$ for
each external line (here $r_1 =4\pi$). 

Therefore, we find for the lowest order boson-boson elastic scattering
\beq
\sigma_{\mu +\mu \ra\mu +\mu}(s)=r_1^4 \frac{\frac{1}{2}(m\Sigma\cos\theta)^2
}{2w^2 (s,\mu^2 ,\mu^2 )} \quad ,\quad r_1 =4\pi
\eeq
which, of course, coincides with a naive computation using the first order
bosonic four-point function $\langle \Phi (x_1)\ldots \Phi (x_4)\rangle_m^c $
(the latter may be inferred immediately from (4)). Observe that (69) 
is singular
at the real particle production threshold $s=4\mu^2$ ($w(4\mu^2 ,\mu^2
,\mu^2)=0$).

In a next step we want to consider the second order contribution of Fig. 8 (the
third type graphs). There are three graphs of this type, namely $s$, $t$ and
$u$ channel, but we will consider only the $s$ channel (annihilation channel)
for the moment. In this diagram the lowest order graph must be subtracted in
order to avoid overcounting (see Fig. 8), therefore the graph of Fig. 12

$$\psannotate{\psboxscaled{600}{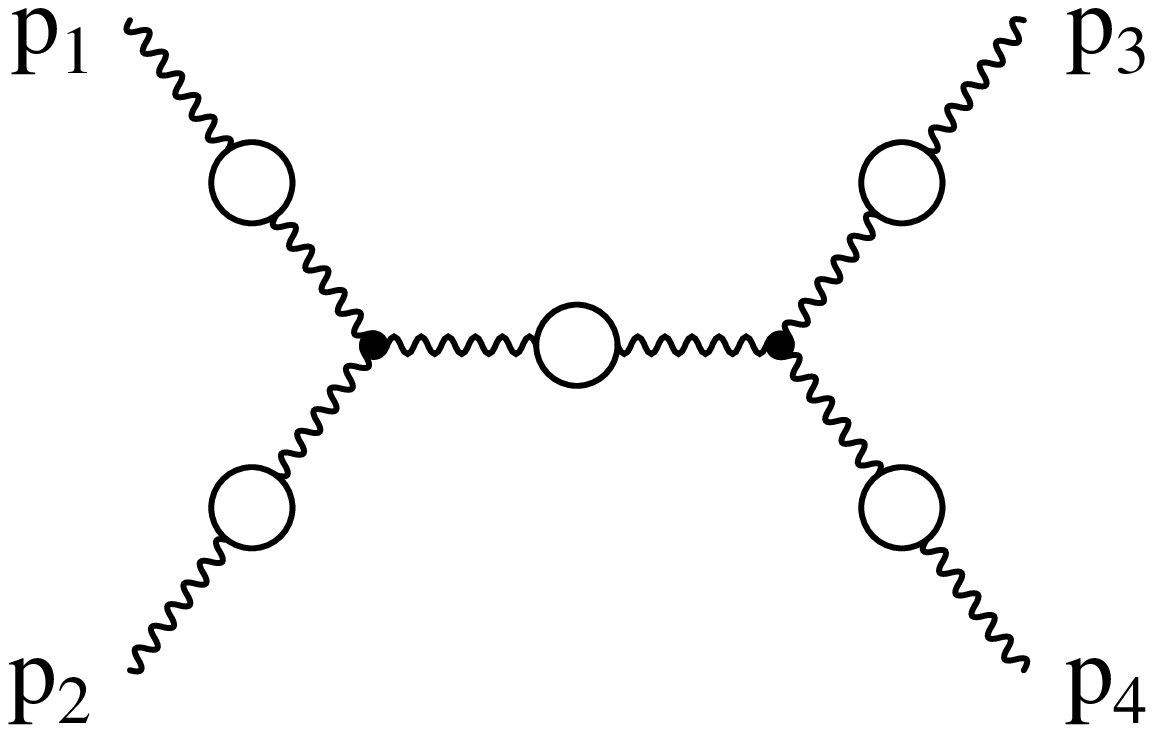}}{\fillinggrid
\at(5.7\pscm;0\pscm){Fig. 12}}$$ 

\vspace{0.1cm}

contains the lowest order graph and the second order $s$ channel contribution.

Actually we will allow for arbitrary final states in the sequel, $\mu +\mu \ra
f$, because this enables us to use the optical theorem, which may be written
for the current problem like
\beq
\sigma^{\rm tot}_{ab\ra f}(s)=\frac{r_a r_b}{w(s,M_a^2 ,M_b^2)}{\rm Im}{\cal
M}_{ab\ra ab}(s)
\eeq
where the $r_i$ are the propagator residues (46) and ${\cal M}_{ab\ra ab}$ is
the forward elastic scattering amplitude. $w(s,M_a^2 ,M_b^2)$ is an initial
state velocity factor; the final state factors must be produced by ${\cal
M}_{ab\ra ab}$, as we will find in the sequel. 

Specifically we choose $a=b=\mu$, and, therefore, both vertices of ${\cal M}$
are contracted by scalars $S$ (we use matrix notation)
\beq
{\cal M}_{2\mu \ra 2\mu}(s)=S^T {\cal G}\Pi (s)S
\eeq
Before starting the computations, we want to make some comments. First, as is
obvious from Fig. 12 and our discussion, in (70) all combinations of $n_1 \mu$
and $n_2 M_2$ are allowed as final states. Consequently, they must exist as
intermediate states in ${\cal M}_{2\mu \ra 2\mu}$, too, in order to saturate
the optical theorem (70). 
Therefore, we are forced to include the $M_2$ particle
into the two-point function $\Pi (p)$, as we did in the previous section, in
order to maintain unitarity.

Secondly, in finite order perturbation theory the optical theorem relates
graphs of different order. However, we use a resummed perturbation series in
(70) and, therefore, we will find a relation that holds for the whole, resummed
two-point function $\Pi (s)$.

In a first step we want to discuss the special case $\theta =0$, because it is
much easier and shows the relevant features without technical complications.
For $\theta =0$ the amplitude (71) reads (see \cite{GBOUND} for an extensive
discussion of the $\theta =0$ case)
\beqa
{\cal M}^{\theta =0}_{2\mu\ra 2\mu}(s) &=&
\frac{m\Sigma}{1-\frac{m\Sigma}{2}(\wt E_+ (s) +\wt E_- (s))} \no \\
&=& \frac{m\Sigma}{1-m\Sigma (d_2 (s) +d_{2,0}(s) +d_4 (s)+\ldots )}
\eeqa
where we inserted the lowest order (30) 
and expanded the exponentials $\wt E_\pm
(s)$ like in (31). Again, we include the $M_2$ particle (which is found by a
further resummation) into $\wt E_\pm$, because this is absolutely necessary, as
we have just argued. Actually $d_{2,0}$ describes the $M_2$-$M_2$ blob, 
and in (72)
only parity even contributions may occur. For the optical theorem (70) we need
the imaginary part
\beq
{\rm Im}{\cal M}^{\theta =0}_{2\mu\ra 2\mu}(s)=\frac{m^2 \Sigma^2 ({\rm Im}d_2
(s) +{\rm Im}d_{2,0}(s)+{\rm Im}d_4 (s)+\ldots )}{[1-m\Sigma ({\rm Re}d_2 (s)+
\ldots )]^2 +m^2 \Sigma^2 ({\rm Im}d_2 (s)+\ldots )^2}
\eeq
We find the following physical picture: at $s=4\mu^2$ the elastic scattering
threshold ($f=2\mu$) opens, at $s=4M_2^2$ the $2\mu \ra 2M_2$ threshold is
added, at $s=16\mu^2$ the $2\mu\ra 4\mu$ threshold, etc. The $d_n (s)$ were
defined as $d_n (s)=\frac{r_1^n}{n!}\wt{D^n_\mu}(s)$, therefore their imaginary
parts are precisely the final state factors for the corresponding cross
section, including the phase space integration (the cutting of the
$\wt{D_\mu^n}(s)$), the propagator normalizations $r_1 =4\pi$, and the final
state symmetry factors for $n$ identical particles, $C_{\rm sym}=\frac{1}{n!}$.
For the multi-$M_2$ propagators $d_{m,0}(s)$ (and, more generally, for
$d_{m,n}(s)$) the first two points (propagators with their residues) are
obvious, the third one (correct $\frac{1}{m!}$ final state symmetry factor) may
be checked by a closer inspection of the mass perturbation series. We show it
for the lowest order contribution to the $M_2$-$M_2$ propagator $d_{2,0}(s)$,
where we depict in Fig. 13 this lowest order contribution and the perturbation
expansion graph where it stems from

$$\psannotate{\psboxscaled{750}{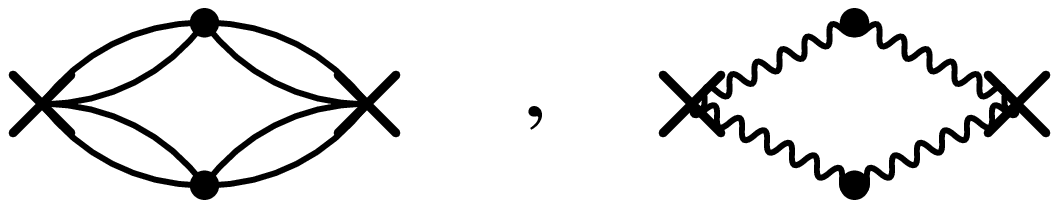}}{\fillinggrid
\at(4.9\pscm;-0.5\pscm){Fig. 13}}$$ 

\vspace{0.1cm}

The second graph in Fig. 13 is a second order mass perturbation, therefore it
contains a factor $\frac{m^2}{2!}$. Further there exists precisely one diagram
of this kind in the perturbation series, therefore the $\frac{1}{2!}$ factor
remains in $d_{2,0}(s)$ as the required final space symmetry factor. Via some
combinatorics this argument may be generalized to higher order contributions to
the $M_2$-$M_2$ loop $d_{2,0}(s)$ and to higher multi-$M_2$ loops.

For the total cross section (70) we get
\beq
\sigma^{{\rm tot},\theta =0}_{2\mu\ra f}(s)=\frac{r_1^2 m^2 \Sigma^2 ({\rm
Im}d_2 (s) +{\rm Im}d_{2,0}(s) +{\rm Im}d_4 (s) +\ldots )}{w(s,\mu^2 ,\mu^2)
\Bigl( [1-m\Sigma ({\rm Re}d_2 (s) +\ldots )]^2 +m^2 \Sigma^2 ({\rm Im}d_2 (s)
+\ldots )^2 \Bigr) }
\eeq
\beq
{\rm Im}d_2 (s)=\frac{r_1^2}{2!}\frac{1}{2w(s,\mu^2 ,\mu^2)} \qquad \mbox{etc.}
\eeq
which we want to evaluate for some specific values of $s$. At the elastic
scattering threshold $s=4\mu^2$, ${\rm Im}d_2 (s)$ is singular and we find
\beq
\sigma^{{\rm tot},\theta=0}_{2\mu\ra f}(4\mu^2)=4.
\eeq
Therefore, the singular behaviour of the lowest order cross section at
$s=4\mu^2$ is cancelled by higher order contributions. This behaviour is,
however, further changed by the $t$ and $u$ channel contributions.

In an intermediate range, far from all thresholds and bound state masses,
$4\mu^2 <s<4M_2^2$, $\sigma^{\rm tot}$ is well described by the lowest order
result (69), because there $m\Sigma d_n (s)$ is small compared to 1,
\beq
\sigma^{{\rm tot},\theta =0}_{2\mu\ra f}(s)\simeq \frac{r_1^2 m^2 \Sigma^2 {\rm
Im}d_2 (s)}{w(s,\mu^2 ,\mu^2)} =\frac{\frac{1}{2!}r_1^4 m^2 \Sigma^2}{2w^2
(s,\mu^2 ,\mu^2)}.
\eeq
At the first bound-state mass, $s=M_{2,0}^2 < 4M_2^2$, a resonance occurs.
There the real part contribution to the denominator of (74) vanishes by
definition and we find
\beq
\sigma^{{\rm tot},\theta=0}_{2\mu\ra f}(M_{2,0}^2)=\frac{r_1^2 m^2 \Sigma^2
{\rm Im}d_2 (M_{2,0}^2)}{w(M_{2,0}^2 ,\mu^2 ,\mu^2)m^2 \Sigma^2 ({\rm Im}d_2
(M_{2,0}^2))^2}=4
\eeq
and the resonance height does not depend on the coupling constant (of course,
the width does).

At the $2M_2$ production threshold $s=4M_2^2$ the scattering cross section goes
down to zero (here $d_{2,0}$ is singular)
\beq
\sigma^{{\rm tot},\theta =0}_{2\mu\ra f}(4M_2^2)\simeq \frac{r_1^2 m^2
\Sigma^2}{w(4M_2^2 ,\mu^2 ,\mu^2)}\frac{{\rm Im}d_{2,0}(4M_{2}^2)}{m^2
\Sigma^2 ({\rm Im}d_{2,0}(4M_2^2))^2}=0.
\eeq
In addition, at this point the $2\mu\ra 2M_2$ production channel opens. At the
four-boson bound-state mass $s=M_4^2$ we find the next resonance
\beq
\sigma^{{\rm tot},\theta =0}_{2\mu\ra f}(M_4^2)=\frac{r_1^2 ({\rm Im}d_2
(M_4^2) +{\rm Im}d_{2,0}(M_4^2))}{w(M_4^2 ,\mu^2 ,\mu^2)({\rm Im}d_2
(M_4^2)+{\rm Im}d_{2,0}(M_4^2))^2}
\eeq
Again, the resonance height does not depend on the coupling constant, and, in
addition, here already two decay channels are open for the $M_4$ resonance.

At the $2\mu\ra 4\mu$ real production threshold $s=16\mu^2$, $\sigma^{\rm tot}$
again vanishes, and for even higher $s$ the above pattern repeats.

Observe that, because $\sigma^{\rm tot}$ has a local maximum (resonance) at
the bound-state masses, whereas it is zero at the real particle production
thresholds, the resonance widths (decay widths) {\em must} be bounded by the
binding energies. For the $M_{1,1}$ and $M_3$ decay widths this may be seen
from the explicit results (56) -- (58).

The $t$ and $u$ channel contributions do not change this pattern (they have no
imaginary parts and are small for all $t$, $u$). 

Next let us turn to the $\theta\ne 0$ case. There parity forbidden transitions
are possible, and therefore we will find $M_{1,1}$ and $M_3$ resonances, too.
The forward scattering amplitude (71) reads
\bdi
{\cal M}_{2\mu\ra 2\mu}(s)= \frac{g_\theta +g^*_\theta -2g_\theta g^*_\theta
(\wt E_+ (s) -\wt E_- (s))}{1-(g_\theta +g^*_\theta )\wt E_+ (s) +g_\theta
g^*_\theta (\wt E_+^2 (s) +\wt E_-^2 (s))} = 
\edi
\beq
\frac{g_\theta +g^*_\theta -4g_\theta g^*_\theta (d_1 (s)+d_{1,1}(s)+d_3
(s) +\ldots )}{1-(g_\theta +g^*_\theta )(d_1 (s)+d_2 (s)+d_{1,1}(s)+\ldots )
+4g_\theta g^*_\theta [d_1 (s)(d_2 (s)+d_{2,0}(s)+\ldots )+\ldots ]}
\eeq
Please observe the presence of only odd $d_i$ in the numerator and of only
odd$\times$even $d_i \times d_j$ in the second term of the denominator.
Therefore, the $4g_\theta g^*_\theta$ terms in the numerator and denominator do
not contribute to parity allowed transitions, and the discussion
of such parity allowed transitions is analogous
to the $\theta =0$ case that we discussed above.

Again, we want to discuss the scattering cross section
\beq
\sigma^{\rm tot}_{2\mu\ra f}(s)=\frac{r_1^2}{w(s,\mu^2 ,\mu^2)}{\rm Im}{\cal
M}_{2\mu\ra 2\mu}(s)
\eeq
for some specific values of $s$. At $s=4\mu^2$ we find again
\beq
\sigma^{\rm tot}_{2\mu\ra f}(4\mu^2)=\frac{r_1^2}{w(4\mu^2 ,\mu^2
,\mu^2)}\frac{(g_\theta +g^*_\theta)^2 {\rm Im}d_2 (4\mu^2)}{1+(g_\theta
+g^*_\theta )^2 ({\rm Im}d_2 (4\mu^2))^2}=4
\eeq
At the first parity forbidden resonance $s=M_{1,1}^2$ we find (${\rm Re}d_{1,1}
=\frac{1}{g_\theta +g^*_\theta}$)
\beqa
\sigma^{\rm tot}_{2\mu\ra f}(M_{1,1}^2) &\simeq & \frac{r_1^2}{w(M_{1,1}^2
,\mu^2 ,\mu^2)}{\rm Im}\frac{g_\theta +g^*_\theta -4g_\theta g^*_\theta {\rm
Re}d_{1,1} (M_{1,1}^2)}{-i(g_\theta +g^*_\theta ){\rm Im}d_2 (M_{1,1}^2)
+4ig_\theta g^*_\theta {\rm Re}d_{1,1}(M_{1,1}^2) {\rm Im}d_2 (M_{1,1}^2)} \no
\\
&=& \frac{r_1^2}{w(M_{1,1}^2 ,\mu^2 ,\mu^2)}\frac{\Bigl( g_\theta +g^*_\theta
-\frac{4g_\theta g^*_\theta}{g_\theta +g^*_\theta}\Bigr)^2 {\rm Im}d_2
(M_{1,1}^2)}{\Bigl( g_\theta +g^*_\theta -\frac{4g_\theta g^*_\theta}{g_\theta
+g^*_\theta}\Bigr)^2 ({\rm Im}d_2 (M_{1,1}^2))^2} \no \\
&=&4
\eeqa
and, therefore, the same resonance height as for the first parity allowed
resonance in the $\theta =0$ case (78).

At the parity forbidden threshold $s=(M_2 +\mu)^2$, where ${\rm Im}d_{1,1}((M_2
+\mu)^2)$ is singular, we find
\bdi
\sigma^{\rm tot}_{2\mu\ra f}((M_2 +\mu)^2) \simeq \frac{r_1^2}{w((M_2 +\mu)^2
,\mu^2 ,\mu^2)}\cdot
\edi
\bdi
\cdot {\rm Im}\frac{g_\theta +g^*_\theta -4ig_\theta g^*_\theta {\rm
Im}d_{1,1}((M_2 +\mu)^2)}{1-i(g_\theta +g^*_\theta )({\rm Im}d_2 
+{\rm Im}d_{1,1}) - 4g_\theta g^*_\theta {\rm Im}d_2 
 {\rm Im}d_{1,1}} =
\edi
\bdi
\frac{r_1^2}{w((M_2 +\mu)^2 ,\mu^2 ,\mu^2)}\frac{(g_\theta +g^*_\theta )^2
({\rm Im}d_2 +{\rm Im}d_{1,1})-4g_\theta  g^*_\theta {\rm Im}d_{1,1}
(1-4g_\theta g^*_\theta {\rm Im}d_2 {\rm Im}d_{1,1})}{(1-4g_\theta g^*_\theta
{\rm Im}d_2 {\rm Im}d_{1,1})^2 +(g_\theta +g^*_\theta )^2 ({\rm Im}d_2 +{\rm
Im}d_{1,1})^2} 
\edi
\beq
\ra \quad  \frac{r_1^2}{w((M_2 +\mu)^2 ,\mu^2 ,\mu^2)}\frac{4g_\theta g^*_\theta
({\rm Im}d_{1,1})^2 {\rm Im}d_2 }{(g_\theta +g^*_\theta )^2 ({\rm Im}d_{1,1})^2
} \quad
\ra \quad \Bigl( \frac{4g_\theta g^*_\theta }{g_\theta +g^*_\theta }\Bigr)^2
\frac{r_1^2 {\rm Im}d_2 ((M_2 +\mu)^2)}{w((M_2 +\mu)^2 ,\mu^2 ,\mu^2)}
\eeq
where we performed the limit ${\rm Im}d_{1,1}\ra\infty$ and kept only the
lowest order contribution in $g_\theta$. Therefore, in contrast to the parity
allowed case, the parity forbidden thresholds do not give zero in $\sigma^{\rm
tot}$.

The reason for this behaviour may be easily understood. In the limit of
$\theta\ra 0$ there should not remain any effect of resonances or thresholds in
$\sigma^{\rm tot}$ for parity forbidden transitions, and $\sigma^{\rm tot}$
should be described by the lowest order result (69). \\
Precisely this happens: Although the resonance height at $M_{1,1}^2$ remains
unchanged for $\theta\ra 0$, (84), its width tends to zero, (56). 
This means that
the resonance $M_{1,1}$ still exists but is stable against  $M_{1,1}\ra 2\mu$
decay for $\theta\ra 0$. Actually the $M_{1,1}$ bound state is a stable
particle at all for $\theta =0$. Further, at threshold $s=(M_2 +\mu)^2$,
$\sigma^{\rm tot}$ tends to the first order result (69) for $\theta\ra 0$,
\beq
\lim_{\theta\to 0} \Bigl(\frac{4g_\theta g^*_\theta }{g_\theta +g^*_\theta
}\Bigr)^2 =m^2 \Sigma^2 +o(m^3)
\eeq 
as it should hold.

Remark: perhaps you would prefer a $\delta$-function like behaviour for the
stable particle limit, $\lim_{\theta\to 0} \sigma^{\rm tot}(M_{1,1}^2)=\infty$,
instead of a constant. Precisely this behaviour is present in the denominator
of (85). 
It behaves like ${\rm Im}\frac{1}{c_{1,1}(s-M_{1,1}^2)-i\epsilon_\theta
{\rm Im}d_2 (s)}$, $\lim_{\theta\to 0} \epsilon_\theta =0$, and, therefore,
leads to the $\delta (s-M_{1,1}^2)$ behaviour for $\theta\to 0$. However, the
whole transition (85) is forbidden for $\theta = 0$. Therefore, the numerator
(residue of the propagator), leads to another zero for $\theta =0$. The two
limits match in a way that leads to a finite result at $s=M_{1,1}^2$, ${\rm
Im}\frac{\epsilon_\theta}{c_{1,1}(s-M_{1,1}^2) -i\epsilon_\theta {\rm Im}d_2
(s)}$. This further shows that the resonance approximation does not give any
contribution at $s\ne M_{1,1}^2$ for $\theta =0$, and, therefore, the lowest
order approximation is the proper one in this case.

For even higher $s$, when both parity allowed and parity forbidden final states
are possible, we again have the problem that the relative sign of the parity
forbidden process is "wrong" due to our conventions (see the remark after equ.
(55)). E.g. at the $M_3$ resonance we find from (81)
\beqa
\sigma^{\rm tot}_{2\mu\ra f}(s=M_3^2) &\simeq & \frac{r_1^2}{w(M_3^2 ,\mu^2
,\mu^2)}\cdot \no \\ 
&&{\rm Im}\frac{g_\theta +g^*_\theta -4g_\theta g^*_\theta {\rm Re}d_3
(M_3^2)}{-i (g_\theta +g^*_\theta )({\rm Im}d_2 (M_3^2)+{\rm Im}d_{1,1}(M_3^2))
+ 4ig_\theta g^*_\theta {\rm Re}d_3 (M_3^2){\rm Im}d_2 (M_3^2)} \no \\
&=& \frac{r_1^2}{w(M_3^2 ,\mu^2 ,\mu^2)}\frac{g_\theta +g^*_\theta
-\frac{4g_\theta g^*_\theta}{g_\theta +g^*_\theta}}{(g_\theta +g^*_\theta
-\frac{4g_\theta g^*_\theta}{g_\theta +g^*_\theta}){\rm Im}d_2 (M_3^2) +
(g_\theta +g^*_\theta ){\rm Im}d_{1,1}(M_3^2)} \no \\
&\simeq & \frac{r_1^2}{w(M_3^2 ,\mu^2 ,\mu^2)}\frac{m\Sigma (\cos\theta
-\frac{1}{\cos\theta})}{m\Sigma (\cos\theta -\frac{1}{\cos\theta}){\rm Im}d_2
(M_3^2) +m\Sigma\cos\theta {\rm Im}d_{1,1}(M_3^2)} \no \\
&\ra & \frac{r_1^2}{w(M_3^2 ,\mu^2
,\mu^2)}\frac{\frac{1}{\cos\theta}-\cos\theta}{(\frac{1}{\cos\theta}-\cos\theta){\rm
Im}d_2 (M_3^2) +\cos\theta {\rm Im}d_{1,1}(M_3^2)} \no \\
&=& \frac{r_1^2}{w(M_3^2 ,\mu^2 ,\mu^2 )}\frac{\sin^2 \theta (\sin^2 \theta
{\rm Im}d_2 (M_3^2) +\cos^2 \theta {\rm Im}d_{1,1}(M_3^2))}{
[\sin^2 \theta
{\rm Im}d_2 (M_3^2) +\cos^2 \theta {\rm Im}d_{1,1}(M_3^2)]^2}
\eeqa
where the last two lines are for real $\theta$ (the last line may be easily
checked by a low order reasoning, in case somebody does not trust our imaginary
$\theta$ convention). 

Again, the resonance height (containing two partial decay channels) does not
depend on the coupling constant.

For even higher $s$ the above pattern repeats.

The last thing to be discussed is the contribution of the $t$ and $u$ channel
diagrams. There the lowest order diagram must be subtracted (see Fig. 8),
\beq
{\cal M'}_{2\mu\ra 2\mu}(t)=S^T{\cal G}(\Pi (t) -{\bf 1})S
\eeq
where $t=4\mu^2 -s\le 0$, $u\equiv 0$.

It is a wellknown fact that the $t$ and $u$ channel amplitudes in the case at
hand have no singularities on the physical sheet of the complex $s$ plane, and,
therefore, no imaginary parts (see e.g. \cite{Eden}). They are, themselves,
imaginary parts of some higher order graphs (in fact, of the non-factorizable
four-point function of Fig. 8). As a consequence, the ${\cal M'}(t)$, ${\cal
M'}(u=0)$ contributions are small for all $t$ and cannot change the
above-discussed behaviour. The only point where ${\cal M'}(t)$, ${\cal M'}(u)$
cause a qualitative change is the elastic threshold $s=4\mu^2$, $t=0$. There
the lowest order singular behaviour, equ (69), that was cancelled by the
$s$-channel contribution, equ. (83), is retained, but with a different
coefficient. We find indeed
\beqa
\sigma_{2\mu\ra 2\mu}(s\sim 4\mu^2) &=& \frac{r_1^4}{w^2 (s\sim 4\mu^2 ,\mu^2
,\mu^2)}|{\cal M}_{2\mu\ra 2\mu}(s\sim 4\mu^2) +2{\cal M'}_{2\mu\ra 2\mu}(0)|^2
\no \\
&\simeq & \frac{r_1^4}{w^2 (s\sim 4\mu^2 ,\mu^2
,\mu^2)}|2{\cal M'}_{2\mu\ra 2\mu}(0)|^2 \no \\
&\simeq & \frac{r_1^4}{w^2 (s\sim 4\mu^2 ,\mu^2
,\mu^2)} 4\Bigl(\frac{2g_\theta g^*_\theta \wt E_- (0) +(g_\theta^2
+g^{*2}_\theta )\wt E_+ (0)}{1-(g_\theta +g^*_\theta )\wt E_+ (0)}\Bigr)^2 \\
&\ra &\infty \qquad \mbox{for} \qquad s\ra 4\mu^2 \no
\eeqa
where ${\cal M}_{2\mu\ra 2\mu} (4\mu^2)$ is given by (71) (including the lowest
order), and ${\cal M'}_{2\mu\ra 2\mu}(0)$ is given by (88) (without lowest
order). The $\wt E_\pm (0)$ are just finite numbers (they have been computed in
\cite{MSSM}).

Further, whenever the $s$-channel cross section vanishes (at parity allowed
higher production thresholds), its value is changed from zero to a small
nonzero number (of order $(m\Sigma)^4)$).

All the other features of the $s$-channel scattering cross section remain
unchanged.

\section{Summary}

We have found the following features of the model in the course of our
discussion:
\begin{enumerate}
\item The Feynman rules of mass perturbation theory acquire a matrix structure
due to the chiral properties of the model (i.e. due to the fact that the
mass term mixes left and right components of fields).
\item Via the Dyson-Schwinger equations of the model all the bosonic $n$-point
functions may be re-expressed in terms of chiral (scalar and pseudoscalar) 
$n$-point functions. Further, these chiral $n$-point functions may be 
re-expressed in terms of non-factorizable ones. These non-factorizable
$n$-point functions are the analogs of the 1PI Green functions in other
theories.
\item For the two-point function this re-expression enabled us to do an
exact resummation of the two-point function $\Pi (s)$ (see (19)). This
resummation made it possible to identify and compute all mass poles of
the stable and instable bound states of the theory, and to find all decay
channels and compute all decay widths for the latter ones.
\item The same re-expression for higher $n$-point functions enabled us to
identify all the possible initial and final states of scattering processes,
and to compute these processes, where the effects of resonances and
higher particle production thresholds are taken into account properly,
without further assumptions or approximations.
\end{enumerate}
As a result, the following physical picture emerged: \\
The model contains two stable particles, namely the Schwinger boson with
mass $\mu$ and the two-boson bound state with mass $M_2$. Further, there
exist (unstable) bound states that are composed of an arbitrary number of
Schwinger bosons and $M_2$ particles. These bound states may decay into all
combinations of $\mu$ and $M_2$ that are possible kinematically. Those bound
states that are composed of some $M_2$ as well are outside the common
knowledge about the massive Schwinger model, but their existence is
enforced by unitarity (by the way, it is not so surprizing, because 
attractive potentials in $d=1+1$ always create at least one bound state).

For scattering processes we found that far from all resonances and particle 
production thresholds the scattering cross section is well described by
a lowest order computation (e.g. the elastic two-particle scattering cross
section behaves like $\frac{1}{s^2}$ for sufficiently large $s$). 
Whenever $s$ is near a bound state, $\sigma (s)$ has a local maximum, i.e. 
a resonance occurs. Moreover, for all values of $s$ where a new final state 
becomes possible kinematically, the corresponding real particle production 
threshold indeed occurs.

All these features are results of the computations, without any further 
approximation in addition to the resummed mass perturbation theory.    

\section {Acknowledgement}

The author thanks R. Jackiw for the opportunity to join the Center of 
Theoretical Physics at MIT, where this work was performed, and the
CTP members for their hospitality.
Further thanks are due to J. Pawlowski and A. Wipf for helpful discussions.

This work is supported by a Schr\"odinger stipendium of the Austrian FWF.

\end{document}